# Heterogeneous Optoelectronic Characteristics of Si Micropillar Arrays Fabricated by Metal-Assisted Chemical Etching

Yang Qian[1], David J. Magginetti[2], Seokmin Jeon[3], Yohan Yoon[3, 4], Tony L. Olsen[5], Maoji Wang[6], Jordan M. Gerton[6], and Heayoung P. Yoon[1, 2]

[1] Electrical and Computer Engineering, University of Utah, Salt Lake City, Utah 84112, USA
[2] Materials Science and Engineering, University of Utah, Salt Lake City, Utah 84112, USA
[3] US Naval Research Laboratory, Washington, DC 20375, USA
[4] Materials Engineering, Korea Aerospace University, Goyang 412-791, South Korea
[5] Utah Nanofab, University of Utah, Salt Lake City, Utah 84112, USA
[6] Physics and Astronomy, University of Utah, Salt Lake City, Utah 84112, USA

Recent progress achieved in metal-assisted chemical etching (MACE) has enabled the production of high-quality micropillar arrays for various optoelectronic applications. Si micropillars produced by MACE often show a porous Si/SiO$_x$ shell on crystalline pillar cores introduced by local electrochemical reactions. In this paper, we report the distinct optoelectronic characteristics of the porous Si/SiO$_x$ shell correlated to their chemical compositions. Local photoluminescent (PL) images obtained with an immersion oil objective lens in confocal microscopy show a red emission peak (≈ 650 nm) along the perimeter of the pillars that is threefold stronger compared to their center. On the basis of our analysis, we find an unexpected PL increase (≈ 540 nm) at the oil/shell interface. We suggest that both PL enhancements are mainly attributed to the porous structures, a similar behavior observed in previous MACE studies. Surface potential maps simultaneously recorded with topography reveal a significantly high surface potential on the sidewalls of MACE-synthesized pillars (+0.5 V), which is restored to the level of planar Si control (-0.5 V) after removing SiO$_x$ in hydrofluoric acid. These distinct optoelectronic characteristics of the Si/SiO$_x$ shell can be beneficial for various sensor architectures.

# 1. INTRODUCTION

Vertically-aligned Si micro/nanopillar arrays have gained tremendous attention for a wide range of applications, including photovoltaic devices [1-4], energy conversion and storage systems [5-7], and chemical/biological sensors [8-10]. Unlike planar geometry, three-dimensional (3D) architecture offers design flexibility, where the optoelectronic characteristics can be optimized by geometrical parameters (*e.g.*, shape, diameter, height, pillar-to-pillar distance). Examples include radial-junction solar cells that allow an optical absorption of sunlight along the length of the pillars (*i.e.*, axial direction), while extracting photocarriers in the radial direction. In our past work, we showed an approximately twofold higher power conversion efficiency with Si micropillar PVs compared to their planar counterparts [3,11,12]. Additionally, the high-aspect-ratio geometry of pillar arrays provides a large surface area per unit substrate, ideally suitable for energy storage and chemical/biological sensors. Functionalized Si nanopillar arrays showed a strong luminescence sensitivity in detecting trace metal ions (*e.g.*, uranyl ions) at a concentration less than 1 ppm [13]. Several research groups also demonstrated various Si nanopillar sensors, where a functionalized pillar surface (*e.g.*, prostate-specific antigen, bio/chemical molecules) detects specific biochemical signals, which are converted to an electrical current [10,14]. Similarly, Si micropillar arrays provide a versatile sensing architecture that consists of a self-aligned active sensing surface (porous shell) and low-resistance electrical interconnectors (crystalline Si core).

To create well-defined pillar arrays, numerous fabrication techniques have been proposed and developed (*e.g.*, deep reactive ion etching [15,16], wafer dicing and cutting [17,18], ion-beam lithography [19,20], metal-assisted chemical etching (MACE) [2,5,21], and metal-catalyzed Vapor-Liquid-Solid [22,23]). Among these methods, MACE is increasingly used for synthesizing Si pillar arrays owing to simple processing, high-throughput, and large-scale production. In MACE, a novel metal film (*e.g.*, Au, Ag, Pt) is used as a catalyst in a mixed etching solution (*e.g.*, $H_2O_2$ and HF in DI water). Theoretical frameworks that describe the detailed electrochemical reactions are still under debate [21,24-26], but below, we summarize two widely accepted models.

Cathode (Au film)

$$H_2O_2 + 2H^+ \xrightarrow{Au} 2H_2O + 2h^+$$
$$2H^+ + 2e^- \rightarrow H_2 \quad (1)$$

Anode (Si substrate)

Model I. 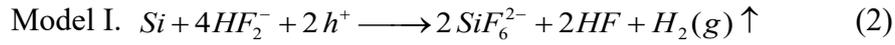
$$Si + 4HF_2^- + 2h^+ \longrightarrow 2SiF_6^{2-} + 2HF + H_2(g)\uparrow \quad (2)$$

Model II. 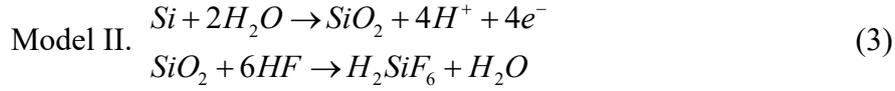
$$Si + 2H_2O \rightarrow SiO_2 + 4H^+ + 4e^-$$
$$SiO_2 + 6HF \rightarrow H_2SiF_6 + H_2O \quad (3)$$

First, the oxidant ($H_2O_2$) near the metal catalyst is reduced, producing holes ($h^+$). The high concentration holes then diffuse to adjacent Si and Si/metal interface, oxidizing Si and generating a soluble compound (hydrofluorosilicic acid; $H_2SiF_6$) and a hydrogen gas ($H_2$) (Equations 2, 3). After the local dissolution of Si, a newly exposed Si surface attracts the metal catalyst via van der Waals interactions and repeats the etching cycles [24]. Model 1 suggests a direct Si dissolution in tetravalent state ($SiF_6^{2-}$; Equation 2), whereas Model 2 proposes two competing processes of oxide formation and Si dissolution (Equation 3) [21,27,28]. Model 2 indicates a presence of $SiO_x$ layer on Si pillars when the relative oxidant concentration ratio of [$H_2O_2$/HF] is sufficiently high, so that the dissolution rate by HF can be slower than that of the oxide formation. High concentration holes can diffuse away from the metal catalyst to adjacent Si if the generated holes ($h^+$) are not fully consumed at the metal-semiconductor interface, introducing structural porosity at the surface of the Si pillar. As a result, MACE processing can produce a porous Si/$SiO_x$ layer on the pillar surface that is composed of Si-rich or pure-Si nanocrystals (< 10 nm in size) mixed in a highly defective SiOx matrix (a few nm to 100's nm thick). Previous studies revealed the complex morphology of this porous Si/$SiO_x$ layer using high-resolution transmission electron microscopy (HRTEM) [2,21,29-32].

Inherent porous structures on the MACE-synthesized pillars can be undesirable for some device applications (e.g., PVs and PN-junction photodetectors) due to a large number of active defects and poor electrical conductivity, and the Si/$SiO_x$ layer is often removed by wet thermal oxidation and stripping processes [33]. On the other hand, porous structures increase the overall surface area, providing additional benefit for sensor systems. Moreover, the porous Si/$SiO_x$ complex has shown remarkably strong luminescence characteristics that can be favorable for a range of optical applications. While most studies have focused on high-aspect-ratio nanopillars (< 1 μm in diameter), an extension of porous Si/$SiO_x$ design to micropillar arrays offers unique optoelectronic configurations, such as an optically active porous surface (shell) and electrically conductive crystalline interconnectors (low-resistance

Si core). However, the critical optical and electrical characteristics of Si micropillar arrays (> 5 μm) have not been thoroughly investigated.

In this paper, we report the optoelectronic characteristics of porous Si/SiO$_x$ shell on Si micropillars fabricated by MACE. To investigate the role of SiO$_x$ complex, we compare the (sub)surface properties before and after removing oxide in a hydrofluoric (HF) solution. A combination of energy-dispersive x-ray spectroscopy (EDS) and numerical simulations is used to measure the compositional variation in a depth-dependent profile. We use confocal microscopy (λ = 405 nm) to visualize the strong red-luminescence along the perimeter of pillars compared to the pillar center. On the basis of our analysis, we illustrate unexpected PL enhancement at the oil/Si shell interface induced by low-fluorescence immersion oil. To measure the surface potential of Si pillar sidewalls, we conduct Kelvin-probe force microscopy (KPFM) on the sidewalls of dispersed Si micropillars. A direct comparison of contact potential difference (CPD) of the pillars to that of planar controls indicates the surface quality of MACE-synthesized Si pillars. Our results confirm the unique characteristics of porous Si/SiO$_x$ shell on micropillar arrays, in turn, supporting versatile sensor design based on MACE-synthesized Si structures.

## 2. RESULTS AND DISCUSSION

MACE is a facile fabrication technique that can produce a variety of 3D structures by simple wet etching processes. Figure 1a shows a schematic of the MACE method, and the inset displays a picture of H$_2$ bubbles arising from the Si substrate in etchant due to the anodic reaction (Equation 2). To test the versatility of the fabrication process, we produce various patterns of arrays, including circles, ribbons, triangles, and squares. The final geometry of the Si arrays depends on processing parameters, including the thickness/shape of metal, the concentration of etchant, and the orientation/doping density of the substrate [2,5]. In this work, we used a 10 nm-thick patterned Au layer as a catalyst. The Au film was thermally-evaporated at a deposition rate of ≈ 0.3 nm/s on a *p*-type (100) Si substrate (resistivity of < 0.005 Ω·cm). The etching solution was prepared by mixing 10 M HF and 0.3 M H$_2$O$_2$ in DI water. The nominal height of the complete Si structures was in the range of 20 μm to 25 μm after the 3-hour etching at room temperature (300 K). Figures 1b ~ 1d show representative scanning electron microscopy (SEM) images of the fabricated Si arrays, closely resembling the intended design. The porous surface on the pillars is often seen in MACE introduced during the local electrochemical reactions (Equations 1~3). To investigate the properties of this porous Si/SiO$_x$ shell, we conduct comprehensive (sub)surface characterizations for the Si micropillar arrays.

Figure 2 shows the compositional distribution of vertically-aligned Si pillar array before (*i.e.*, pristine pillar array) and after removing SiO$_x$ complex in hydrofluoric acid (5M HF for 1 minute). The energy-dispersive x-ray spectroscopy (EDS) datasets were collected with an electron-beam (e-beam) rastering over a small area (≈ 40 μm × 50 μm) on top of the pillar arrays. Two prominent x-ray signals at 0.5 keV and 1.7 keV are corresponding to oxygen (O $K_α$) and silicon (Si $K_α$), respectively (Figure 2a). The integrated intensity ratio of [Si $K_α$]/ [O $K_α$] increased from 2 to 40 after the HF treatment, indicating a significant amount of oxide was present on the pillar surface after MACE processing. The residual amount of oxygen in the HF treated pillar array is likely attributed to a native oxide formed in air ambient.

We collected a series of EDS spectra at different acceleration beam voltages ($V_{acc}$) to obtain a depth-dependent compositional profile. Since an e-beam travels deeper into the sample at a higher $V_{acc}$, the keV-dependent EDS can provide qualitative chemical information from surface to bulk interior of Si micropillar arrays. Using Monte-Carlo simulations [34,35], we estimate the e-beam probe depth ($d_p$) of 220 nm at 5 keV, 800 nm at 10 keV, 1500 nm at 15 keV, and 2700 nm at 20 keV for Si, where the injected e-beam loses 90 % of its initial energy within the corresponding interaction bulb. An example of the full scale of energy loss contours under a 5 keV e-beam irradiation is shown in Figure 2c, while displaying the simulated Si $K_α$ profile at 10 keV in Figure 2d. Considering the overvoltage that needs to be two- to three-fold higher than the characteristic x-ray signal (*i.e.*, 1.7 keV Si $K_α$ in this study) [36,37], we set a minimum $V_{acc}$ of 5 keV for the keV-dependent EDS. After applying a ZAF correction (Z: atomic number effects, A: absorption, F: fluorescence), the mass fractions of O and Si were extracted and plotted in Figure 2b. For the pristine Si pillar array, the oxygen mass fraction of ≈ 40 % at 5 keV ($d_p$ = 220 nm) gradually decreases to ≈ 25 % at 20 keV ($d_p$ = 2700 nm), indicating an oxygen-rich Si pillar surface. After the HF treatment, the mass fraction of oxygen becomes negligible, whereas the Si fraction is approaching unity. Our EDS analysis suggests that a substantial amount of SiO$_x$ is present on the Si pillar array, which is likely introduced during the MACE processing (Model 2).

Photoluminescence (PL) spectroscopy is frequently used to identify optically active defects by revealing their energy within the energy band-gap. Unlike Si bulk semiconductors (indirect band-gap of $E_g$ = 1.12 eV), which show no noticeable PL emission, extensive studies have reported strong luminescent characteristics of porous Si/SiO$_x$ structures. We use confocal PL microscopy to map the spatially-resolved PL emission of a porous Si/SiO$_x$ shell on Si micropillar arrays. Figure 3a illustrates a simplified PL setup used in this work. A laser beam (λ = 405 nm, power = 1.5 mW) is directed through a high numerical aperture (NA) objective via inverted microscope system, and the emitted PL signals are collected through the same objective lens and detected by a charged coupled device (CCD) camera/spectrometer. Details of the experimental setup can be found elsewhere [38]. Figure 3b shows a typical PL spectrum (400 nm to 880 nm) obtained from our Si micropillar arrays using an air objective lens (40×, NA = 0.6). We observe a pronounced luminescence peak near 670 nm. A Gaussian peak yields 1.9 eV for the energy of the peak and a full-width half max (FWHM) of 0.3 eV, which is consistent with the literature [29,39,40]. Prior studies suggest that the origin of this red band is associated with the porous Si/SiO$_x$ layer, where the local electrochemical etching creates Si nanocrystals (< 10 nm in size) [29,41-43] as well as active defect sites, such as non-bridging oxygen hole centers (NBOHC, ≡Si-O•). The emission peak for NBOHC is centered around 1.9 eV [40,41], whereas the luminescence of Si nanocrystals is in the range of ultraviolet (3.5 eV) to infrared (1.5 eV), depending on the cross-sectional area of the Si nanocrystals [44]. The prominent red emission of the pillar array can be a convolution of these contributions.

Next, we consider the local PL characteristics from individual pillars (*e.g.*, perimeter vs. center of pillars) using high-resolution PL imaging with an oil immersion objective lens (100×, NA=1.4). A vertically-aligned pillar array specimen was placed on a glass coverslip, and the air gap between them (*i.e.*, sample, coverslip, and objective lens) was filled with commercially available immersion oil (Cargille Type B) to improve spatial resolution by matching the refractive indices (Figure 3a). We estimate a PL sampling volume of < (300 nm)$^3$ in our setup based on the diffraction limit and our confocal microscopy configuration [38,45]. Prior to the PL imaging of Si pillar arrays, we measured the background luminescence of the stack of layers of oil/coverslip/oil/objective lens without the pillar specimen. Figure 3c displays the PL spectrum collected for a long integration time (10 s), showing a broadband luminescence centered around 520 nm (≈ 2.4 eV). The sharp short peak on the shoulder at 460 nm (≈ 2.7 eV) is attributed to the optical components of our setup rather than a sample signature. We subtract this constant background signal from the PL datasets of the pillar array based on multiple Gaussian curve fitting. Figure 3d displays a panchromatic PL image collected on top of the vertically-aligned Si micropillar array. A focused laser beam (405 nm) was scanned in the area of 35 μm × 35 μm (64 pixels × 64 pixels) at 0.03 Hz (≈ 208 ms per pixel), and each pixel records the corresponding PL spectrum from 400 nm to 880 nm. The brightness contrast in the PL image reflects the changes of the integrated PL intensity across all wavelengths. We find the strongest PL emission along the perimeter of each pillar.

To investigate the contribution of each PL band, we extract a series of PL spectra along the line across a pillar diameter, as indicated by the yellow line in Figure 3d. The PL plots (Figure 3e) reveal the expected red and blue band at ≈ 650 nm and ≈ 520 nm, respectively. As observed in the control experiment (Figure 3c), the shallow peak at ≈ 460 nm is attributed to our PL setup, remaining approximately unchanged for all spectra. Interestingly, both red and blue bands show the highest emission intensity near the perimeter of the pillars and rapidly decrease at the center and outside of the pillar. The trends of the peak position, height, and FWHM for each band were further quantified using multiple Gaussian curve fitting. Figures 3f and 3g show a summary of the results (see the Supporting Information for curve fitting of the individual spectrum). The peak positions of the red band (650 nm; 1.9 eV; Peak 1) and the blue band (520 nm; 2.4 eV; Peak 2) are relatively uniform across the pillar diameter, with a slight increase toward the pillar center (Δ < 0.05 eV) and outside the pillars (Δ < 0.09 eV). The FWHM of the red band estimated from the individual spectra is in the range of 0.4 eV ~ 0.5 eV, slightly broader than the peak measured with a 40× air objective (≈ 0.3 eV; Figure 3b). The overall variations of FWHM for both red and blue emissions are also marginal (Δ < 0.1 eV) within the pillar. The 460 nm peak (≈ 2.7 eV) shows a negligible FWHM deviation. On the basis of the quantitative comparison, we find that the red emission at 1.9 eV is significantly stronger (> 3×) occurs near the perimeter of the pillar (*i.e.*, shell) compared to the center and outside the pillar (Figure 3f). The higher PL intensity indicates an increase of Si/SiO$_x$ structural porosity, which also enhances the radiative recombination centers, such as non-bridging oxygen hole centers (NBOHC) around 1.9 eV. Similar PL behaviors have been reported for high-aspect-ratio Si nanopillars, where the PL intensity is the brightest near the highly porous tip (< 2 μm along the axial direction) [39]. Peak 2 (520 nm; 2.4 eV) also shows a similar trend seen in Peak1, about 2× stronger luminescence at the perimeter relative to the center of the pillar. The PL spectra collected outside of the pillar (*i.e.*, pixel ID of #1 and #11 in Figure 3e) are nearly identical to the control PL (*i.e.*, oil/glass/oil/objective lens without a pillar specimen; Figure 3c), suggesting that the blue PL is associated with the fluorescence of the immersion oil. However, unlike the red emission, this apparent PL enhancement at 2.4 eV cannot be explained by the increase of $SiO_x$ defect centers (1.9 eV). We speculate that the strong blue PL emission could be associated with the porous Si structures, as proposed in the quantum confinement/luminescent center model (QCLC) by Qin *et al.* [43]. In this model, the Si nanocrystals primarily serve as a reservoir for exciton production via quantum confinement. Under an external excitation (*e.g.*, 405 nm [≈ 3.1 eV] laser beam in our case), Si nanocrystals in the porous $Si/SiO_x$ shell generate excitons that can tunnel through to the neighboring luminescent centers and are radiatively recombined. This QCLC model also supports the underlying mechanism for the remarkably strong (> 3×) red luminescence observed on the $Si/SiO_x$ shell [29].

A key advantage of MACE over established plasma-based techniques (*e.g.*, deep reactive ion etching) is to control the etching process close proximity with a metal catalyst (*e.g.*, a few 10's nm), so that possible structural damage (*e.g.*, porosity, defects) remains only at the surface, while preserving the initial high-quality in the core of the Si pillars. Our EDS analysis shows a strong oxygen peak (Figure 2), indicating that the surface electronic states could be altered during the local oxidation and dissolution. To address this possibility, we perform KPFM on the sidewall of micropillars that directly measures the surface potential at the nanoscale (Figure 4). Atomic force microscopy (AFM) images were simultaneously collected with KPFM to correlate electronic structure to their morphology. We prepared two specimens by cleaving a Si micropillar array (as-synthesized) into two pieces. One of them was dipped into a 5 M HF solution to remove the $SiO_x$ layer. Each pillar array was then sonicated in a vial containing IPA and dispersed on an Au patterned Si substrate. The SEM image in Figure 4 (a) shows the dispersed individual micropillars. The inset of the figure illustrates the KPFM configuration, where a KPFM tip is in contact with the sidewall of a Si pillar. AFM images of Si pillars before and after oxide removal are shown in Figures 4c-4f. The large striations are likely associated with the lithography process to pattern the Au film, whereas the porous $Si/SiO_x$ structures are attributed to the local electrochemical etching. We used the polynomial background subtraction to calculate the overall surface roughness in root-mean-square (RMS), resulting in 15.6 nm for the pristine (*i.e.*, as-synthesized by MACE) pillars and 10.8 nm for the HF etched pillars. Figure 4c and 4d show the representative AFM linescans, where the peak-to-valley roughness notably decreases with the oxide removal in HF, from approximately 12 nm to 7 nm, which reveals the shape of Si nanostructures on the pillar surface. These results also suggest that the local electrochemical oxidation preferentially occurs at protruding features on the Si pillar due to kinetic limitation, forming thicker $SiO_x$ than on smooth regions, which is often seen in conventional wet or dry oxidation of Si [46]. Upon HF treatment, the $SiO_x$ layer is removed, effectively smoothing out the surface. We note a native oxide can easily form on the H-terminated Si surface (Si-H) after the HF treatment due to the exposure to ambient conditions.

KPFM images collected on the same area for the AFM are shown in Figure 4 (g, h). Each pixel in the KPFM maps records the contact potential difference (CPD; $V_{CPD}$) that is defined as the work function difference between the sample ($\phi_{sample}$) and the probe tip ($\phi_{tip}$) [47-49], where $e$ is the elementary charge ($1.6 \times 10^{19}$ C).

$$V_{CPD} = \frac{\phi_{tip} - \phi_{sample}}{-e} \quad (4)$$

There are several bright-line contrasts in the KPFM images, particularly with the pristine pillar (Figure 4g). It is well recognized that abrupt change in topography can influence KPFM signals, or it could be related to accumulated local charges. Considering that the bright lines in the KPFM image follow the same morphology observed in the AFM (Figure 4 e, f), the high surface potential is likely attributed to an artifact associated with the surface morphology of the samples. The overall distribution of CPD in the KPFM images shows very little dependence on the pillars' curvature. For quantitative analysis, we constructed the CPD line scans for each sample and compared to planar controls (Figure 4b). A representative CPD line scan for pillars was extracted from each potential map (Figures 4 g, h), while collecting an equivalent length of CPD for planar controls (Au film on Si, bare Si substrate). The overall CPD magnitude of the control samples remains constant, about -0.5 V (planar Si) and -0.7 V (Au film). The average CPD value of the as-synthesized Si pillar reaches to ≈ 0.5 V, approximately 1 V higher than that of Si control. Interestingly, this high CPD value decreases back to -0.5 V after the HF treatment. We note that a native oxide is present in all our samples, even for the HF-treated pillars, due to the exposure to air ambient prior to the KPFM measurements. Therefore, the high surface potential of the as-synthesized pillar is likely attributed to the $SiO_x$ complex that was produced during the MACE process. It is also possible that the isolated Si nanocrystals embedded in a $SiO_x$ matrix influence the high surface potential. The dissolution of the $SiO_x$ matrix in HF exposes the underlying Si surface, which has similar qualities to that of Si planar control. The porous Si structure could have inhomogeneous local electronic structures, possibly leading to a large dispersion of the CPD values (≈ 0.6 V peak-to-peak).

We determined the work function of our samples by calibrating the KPFM probe tip (Table 1). Highly-oriented pyrolytic graphite (HOPG) was used as a calibration sample, which has a known work function of 4.6 eV [50]. To ensure the same tip condition for each KPFM scan, we also used the patterned Au film as a reference (work function of 5.1 eV), where the Si micropillars were dispersed (Figure 4a). Our Si micropillars were fabricated with a highly-doped *p*-type substrate (resistivity < 0.005 Ω·cm), and the corresponding Fermi-level was located close to the valence band ($E_v$). The measured work function of the Si planar sample shows a good agreement with the estimated value of ≈ 5 eV. Table 1 summarizes the calculated work function of each sample using Equation (4).

## 3. CONCLUSIONS

In summary, we show subsurface optoelectronic characteristics of Si micropillar arrays fabricated by metal-assisted chemical etching. To examine the Si/SiO$_x$ complex introduced during the MACE synthesis, we compare the surface properties of the Si pillars before and after removing SiO$_x$ in a hydrofluoric acid (HF) solution. Confocal PL images confirm that the perimeter of the pillars exhibits strong PL signals (3×) compared to the pillar center. In addition, we find a notable PL increase (≈ 540 nm) at the oil/shell interface. We speculate that the porous Si/SiO$_x$ shell of the pillars is responsible for these optical enhancements, a similar behavior reported in prior literature. We measure contact potential difference (CPD) of pillar sidewalls and planar controls, which demonstrates that a notably different surface potential with as-synthesized pillars. The high surface potential of MACE-synthesized pillars is restored to the level of planar Si control after removing SiO$_x$ in HF. Our findings support that the distinct optoelectronic characteristics of the Si/SiO$_x$ shell can be beneficial for sensors consisting of self-aligned active sensing surface (shell) and low-resistance electrical interconnectors (core).

## 4. MATERIALS AND METHODS

*a. Fabrication of Si Micropillar Arrays*

Vertically-aligned micropillar arrays were fabricated with (100) *p*-type Si wafers (resistivity < 0.005 Ω·cm) via conventional lithography techniques [4]. Briefly, the wafers were cleaned in a series of solvents (acetone, isopropanol, and DI water) and blown dry with nitrogen gas ($N_2$). Following a double-layer photoresist coating (LOR 10B, Microchem 1813) and a soft-bake process, the samples were patterned with arrays of micropillars. An $O_2$ plasma descum (≈ 50 W for 30 s) was performed to remove organic photoresist residue. A thin layer of Au film was evaporated (thickness: ≈ 30 nm; deposition rate: ≈ 0.3 nm/s) and lifted off. We performed an $O_2$ plasma descum (≈ 100 W for 60 s) prior to metal-assisted chemical etching (MACE). Figure 1a illustrates the MACE processing. The patterned Si wafer was immersed in a mixed chemical solution (10 M HF and 0.3 M $H_2O_2$ in DI water) at room temperature for about 3 hours. The bubble formation in the inset of Figure 1a was observed, confirming the production of $H_2$ gas, as shown in Equation (1). After the MACE process, the residual Au film was removed in an Au etchant, and the Si pillar arrays were cleaned in solvent and blown dry with $N_2$. Figure 1c shows a portion of the completed Si pillar arrays.

*b. Energy-Dispersive X-ray Spectroscopy*

Energy-dispersive x-ray spectroscopy (EDS) was performed on top of vertically-aligned Si pillar arrays before and after removing $SiO_x$ complex in hydrofluoric acid (5M HF for 1 minute). The EDS detector was equipped in a scanning electron microscopy (SEM) system, where the primary electron beam voltage can be changed from 1 keV to 30 keV. Quantitative analysis was performed using a ZAF correction (Z: atomic number effects, A: absorption, F: fluorescence) and background subtraction in a commercially available software (EDAX TEAM).

*c. Photoluminescent Characterization*s

PL characterizations were performed in a confocal PL microscopy system. A 405 nm laser beam (power ≈ 1.5 mW) was illuminated through a high numerical aperture (NA) objective via inverted microscope system, and the emitted PL signals are collected through the same objective lens and detected by a charged coupled device (CCD) camera/spectrometer (Figure 3a). We used an air objective lens (40×; NA = 0.6) to collect the baseline PL spectra of Si pillar arrays, and an oil-objective lens (100×; NA = 1.4) was used for PL imaging.

To form a PL image, we scanned a sample area of 35 μm × 35 μm (64 pixels × 64 pixels) at a scan rate of 0.03 Hz. We estimate a sampling time of ≈ 208 ms per pixel by considering of the trace, retrace, and turn-around time of the piezoelectric actuator. A spectrum was collected at each pixel in a wavelength range of 400 nm to 880 nm.

*d. Kelvin-probe force microscopy (KPFM)*

Prior to the AFM/KPFM, we prepared two sets of samples. A Si micropillar array (as-synthesized) was cleaved into two pieces, and one of the samples was dipped into a 5 M HF solution for 1 minute to remove the $SiO_x$ layer. Each pillar array sample was then sonicated in a vial containing IPA and dispersed on an Au patterned Si substrate (Figure 4a). AFM/KPFM characterizations were conducted using a scanning probe microscopy system (Bruker Dimension Icon SPM) equipped with a doped Si probe (PFQNE-AL, Bruker) that has a nominal tip radius of 5 nm and a nominal spring constant of 1.5 N/m. We used a frequency modulation mode (FM-KPFM, $f_0 \approx 262$ kHz). Topography and contact potential difference (CPD) of the samples were obtained simultaneously in a single-pass tapping mode [49].

## ACKNOWLEDGMENTS

The authors thank D. Segovia, K. Powell, B. Baker, P. Perez, and B. Van Devener for valuable discussions and experimental assistance during this work. This research was supported by a University of Utah Seed Grant and New Faculty Start-up Funds. We acknowledge support by the USTAR shared facilities at the University of Utah, in part, by the MRSEC Program of NSF under Award No. DMR-1121252. The KPFM experiments were performed while S. Jeon and Y. Yoon held an NRC Research Associateship and an ASEE Postdoctoral Fellowship at the U.S. Naval Research Laboratory, respectively.

## AUTHOR CONTRIBUTIONS

YQ and DJM performed the experiments and analyzed the datasets. SJ and YY performed the KPFM characterizations. MJ helped the PL data acquisition, and JMG supervised the PL measurements. TLO and HPY developed the fabrication processes. HPY supervised the project. All authors contributed to manuscript revision, read and approved the submitted version.

## CONFLICT OF INTEREST

The authors declare no conflict of interest.

**Figure 1.**

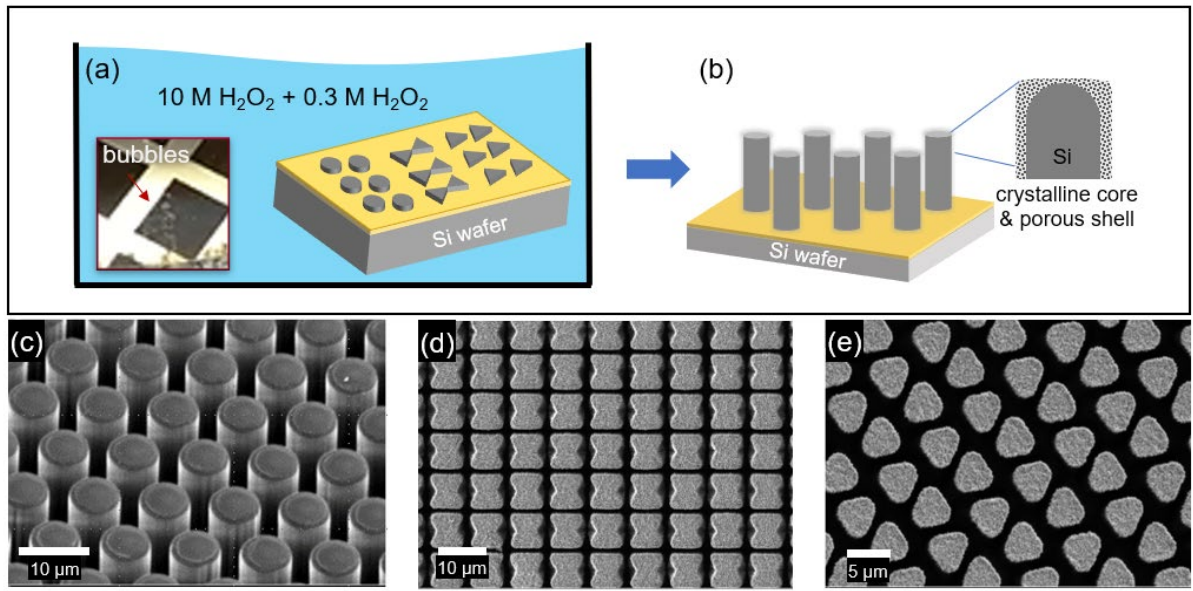

Schematics of (a) and (b) illustrate the metal-assisted chemical etching (MACE) processes to fabricate 3D Si micropillar arrays. The $H_2$ bubbles arise from the Si substrate in the etchant due to the anodic reaction (Equation 2). Scanning electron microscope (SEM) image of a portion of the MACE fabricated 3D structures with an array pattern of circles (c), ribbons (d), and triangles (e). Comprehensive optoelectronic characterizations were performed on the pillar arrays in (c) that have a nominal diameter of 8 μm and a height of 20 μm, and the distance between the pillars is about 3.5 μm.

**Figure 2.**

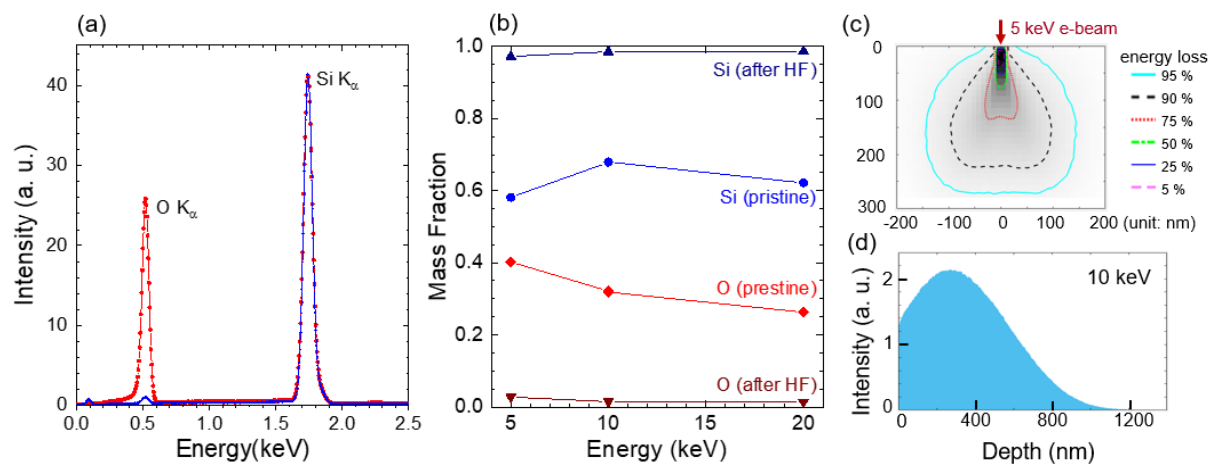

(a) Energy-dispersive x-ray spectroscopy (EDS) plots obtained on the Si micropillar arrays before and after removing $SiO_x$ in HF. Two prominent x-ray signals at 0.5 keV and 1.7 keV correspond to oxygen (O $K_\alpha$) and silicon (Si $K_\alpha$). (b) The mass fraction of the pillars after applying a ZAF correction (Z: atomic number effects, A: absorption, F: fluorescence). Examples of Monte-Carlo simulations showing a full scale of energy loss contours under a 5 keV e-beam irradiation (c) and Si $K_\alpha$ profile at 10 keV (d). An estimated e-beam probe depth by simulations is 220 nm at 5 keV, 800 nm at 10 keV, 1500 nm at 15 keV, and 2700 nm at 20 keV for Si, respectively, where the injected e-beam loses the 90 % of initial energy within the corresponding interaction bulb.

**Figure 3.**

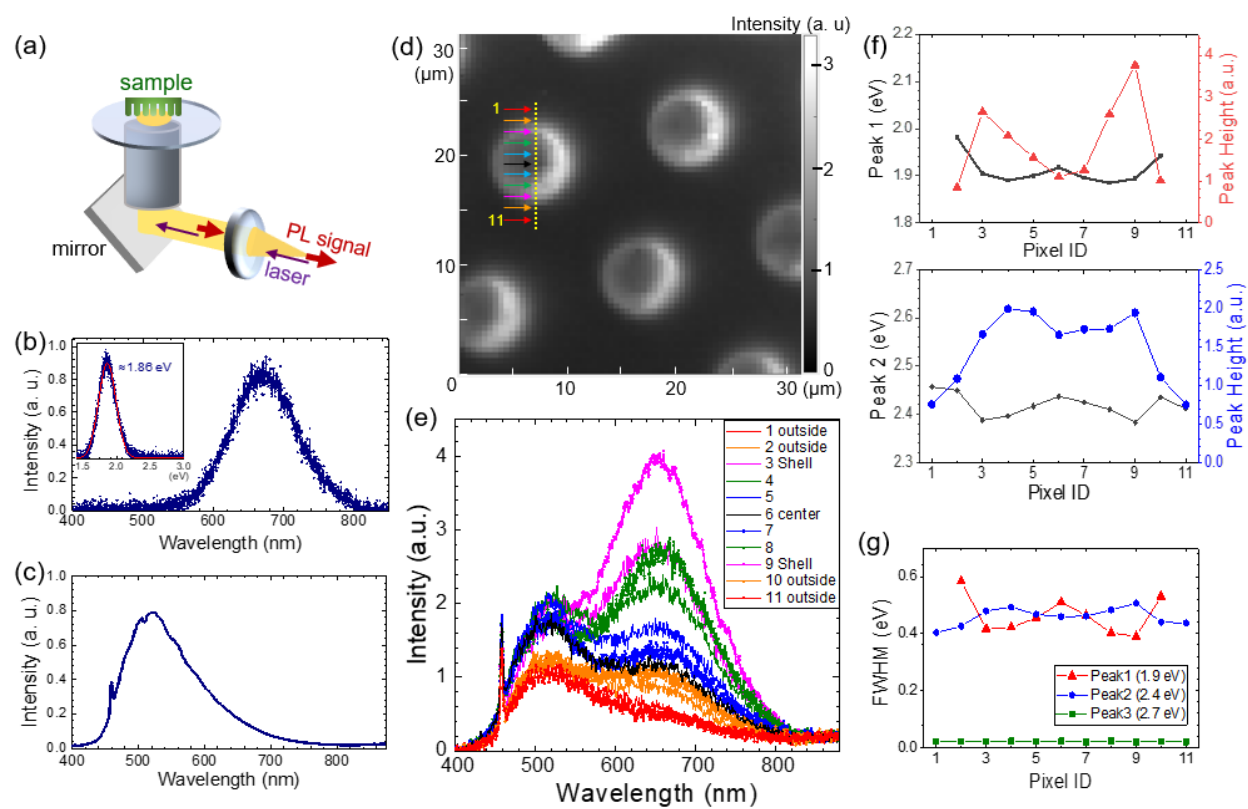

(a) A schematic of inverted confocal photoluminescence (PL) set-up, where a laser beam (λ = 405 nm, power = 1.5 mW) is illuminated through a high numerical aperture (NA) objective, and the emitted PL signals are collected through the same objective lens and detected by a charge-coupled device (CCD) camera/spectrometer. (b) A PL spectrum of vertically-aligned Si pillar array obtained with an air objective lens (40×; NA = 0.6). (c) Background PL luminescence (10 s integration time) induced by the stack of layers of oil/coverslip/oil/objective lens without the pillar specimen. (d) A high-resolution panchromatic PL image (400 nm ~ 880 nm) collected with an oil immersion objective lens (100×; NA = 1.4). (e) A series of PL linescans extracted from the yellow line on the image. (f) and (g) show the extracted peak information from the multiple Gaussian curve fitting of the PL linescans: Peak 1 (650 nm; 1.9 eV), Peak 2 (520 nm; 2.4 eV) and Peak 3 (460 nm; 2.7 eV).

**Figure 4.**

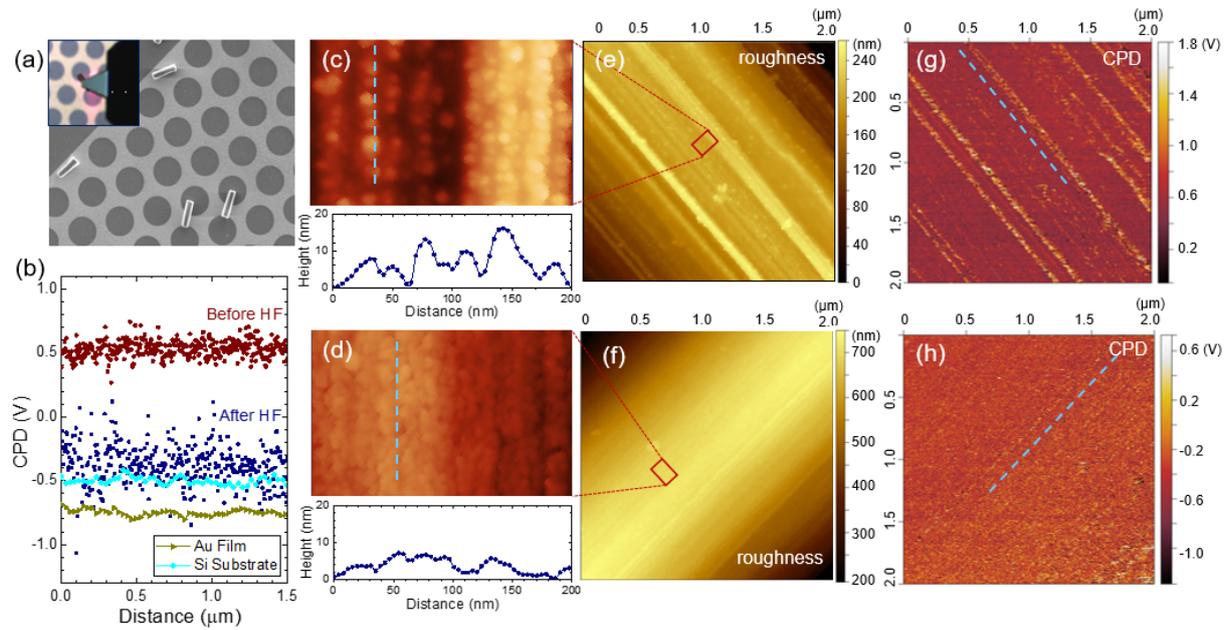

(a) Scanning electron microscopy (SEM) image of the dispersed Si micropillars on a patterned Au film on Si. The inset illustrates the Kelvin-probe force microscopy (KPFM) configuration, where a KPFM probe is scanning the sidewall of a Si pillar (b) Contact potential difference (CPD) obtained on the Si pillars and the planar controls (Au film, Si substrate). Atomic force microscopy (AFM)/KPFM performed on the Si pillar sidewalls before (c, e, g) and after (d, f, h) removing $SiO_x$ in HF.

**Table 1**. The calculated work function of the Si pillars before and after oxide removal in hydrofluoric acid (HF). Mean and standard deviation (in parenthesis) of the contact potential difference (CPD) are shown in the column of surface potential. Highly-oriented pyrolytic graphite (HOPG) was used as a calibration sample with a known work function of 4.6 eV [50]. The patterned Au film as a reference (work function of 5.1 eV), where the Si micropillars were dispersed (Figure 4a).

| | Sample | Surface potential (mV) | Sample work function (Ref. of HOPG) (V) | Sample work function (Ref. of Au) (V) |
|---|---|---|---|---|
| Control | Cleaved HOPG | -360 (50) | Ref. | 4.83 |
| | Au film on Si substrate | -734 (56) | 4.97 | Ref. |
| | Si substrate | -506 (58) | 4.75 | 4.87 |
| As-synthesized | Si pillar sidewall | 611 (212) | 3.63 | 3.97 |
| | Au film on Si substrate | -515 (93) | 4.76 | Ref. |
| | Si substrate | -187 (83) | 4.43 | 4.77 |
| HF treated (oxide removal) | Si pillar sidewall | -352 (241) | 4.59 | 4.82 |
| | Au film on Si substrate | -636 (52) | 4.88 | Ref. |
| | Si substrate | -510 (55) | 4.75 | 4.97 |

**Supporting Information**

We use multiple Gaussian curve fit functions to decompose the PL peaks. The same curve fit was applied for eleven PL spectra in Figure 3e. Three peaks around 1.9 eV, 2.4 eV, and 2.7 eV were observed for all individual spectrum. The extracted data were plotted in Figure 3f and 3g.

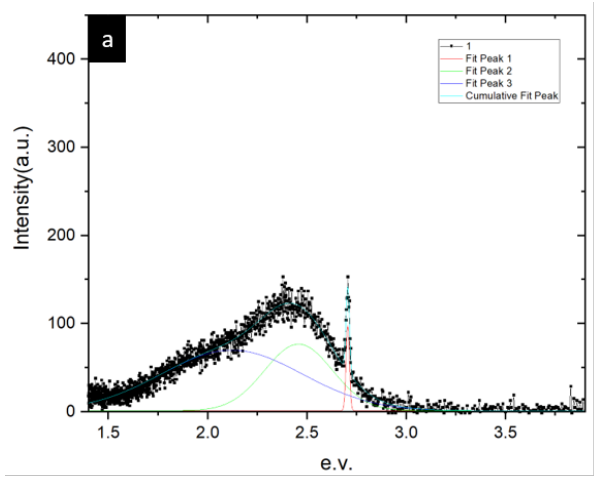
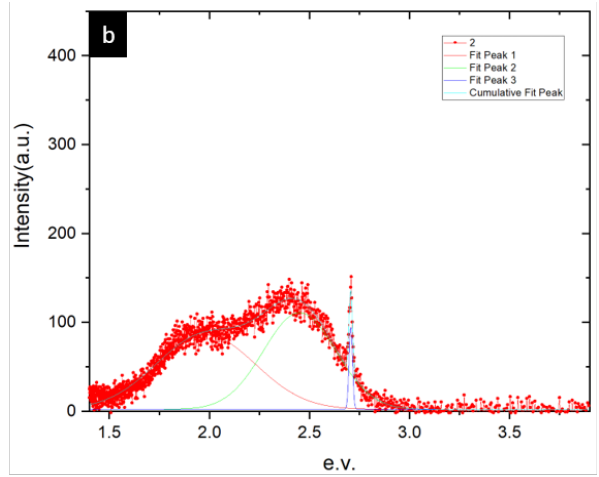
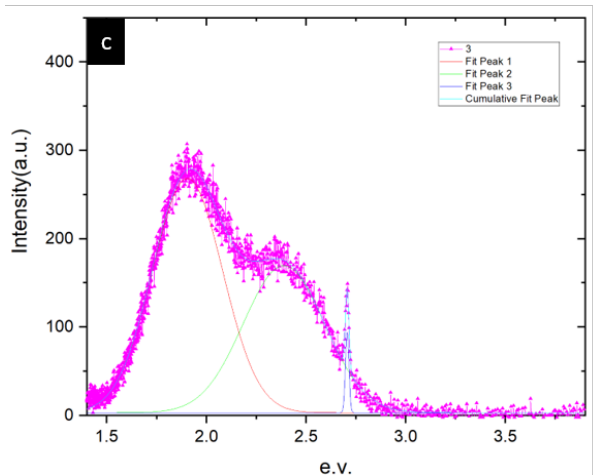
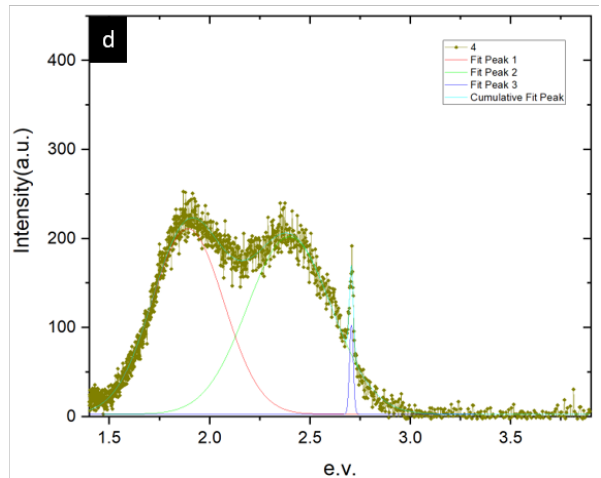

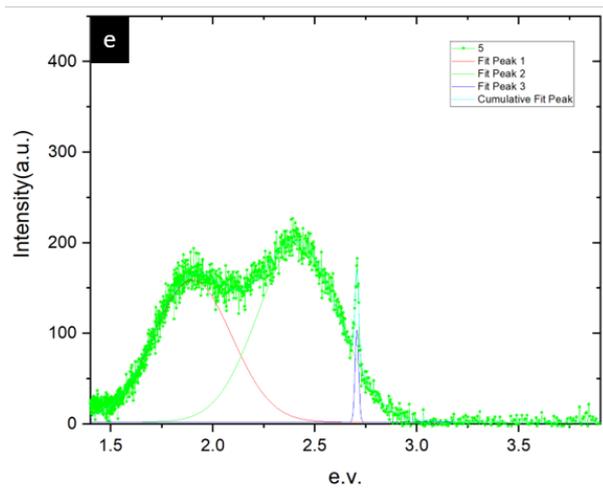
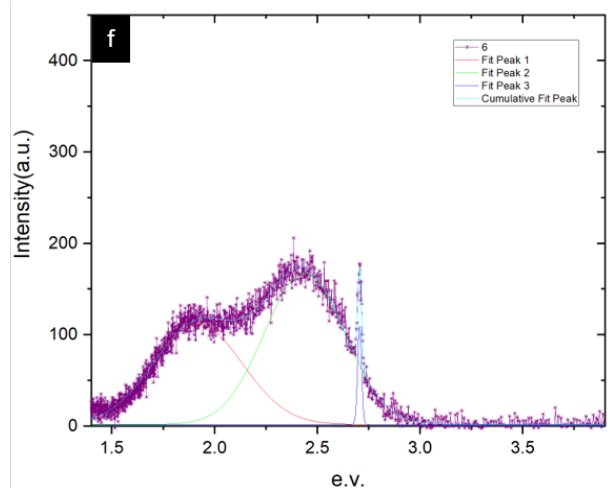
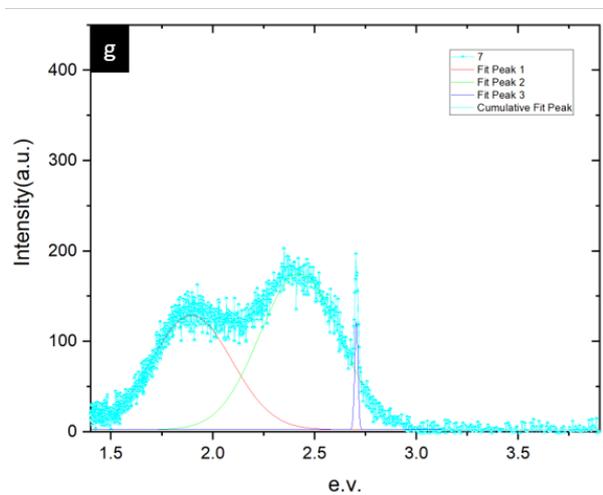
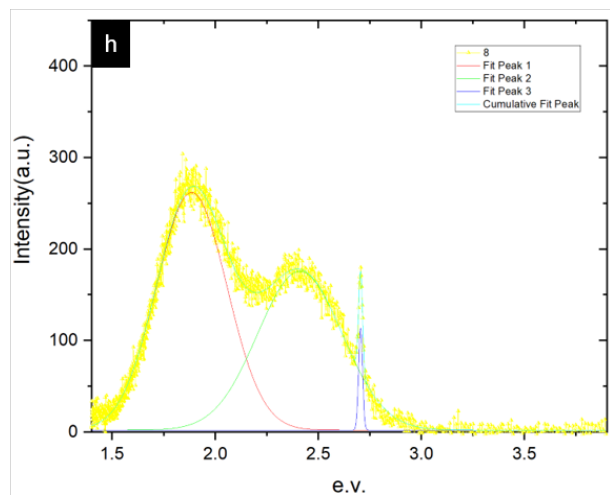

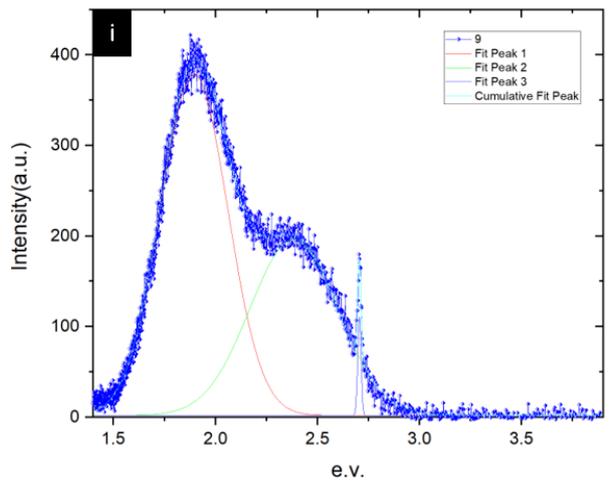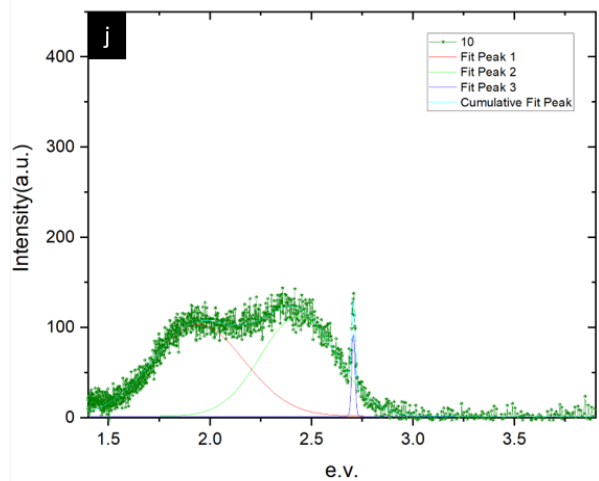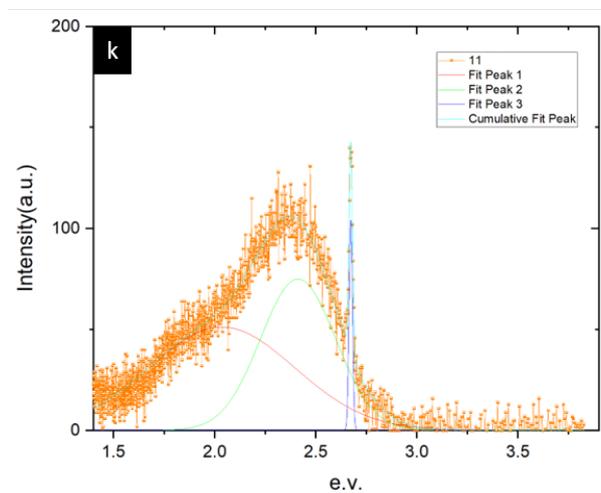